# Lone Pair Induced 1D Character and Weak Cation-anion Interactions: Two Ingredients for Low Thermal Conductivity in Mixed-anion Metal Chalcohalides

Xingchen Shen,[1*] Koushik Pal,[2] Paribesh Acharyya,[1] Bernard Raveau,[1] Philippe Boullay,[1] Carmelo Prestipino,[1] Susumu Fujii,[3] Chun-Chuen Yang,[4] I-Yu Tsao,[5] Adèle Renaud,[6] Pierric Lemoine,[7] Christophe Candolfi,[7] Emmanuel Guilmeau[1*]

[1]CRISMAT, CNRS, Normandie Univ, ENSICAEN, UNICAEN, 14000 Caen, France

[2]Dept. of Physics, Indian Institute of Technology Kanpur, Kanpur, UP 208016, India

[3]Department of Materials, Faculty of Engineering, Kyushu University, Fukuoka 819-0395, Japan

[4]Department of Physics, National Central University, Chung-Li District, Taoyuan City, 320317, Taiwan

[5]Institute of Materials Science and Engineering, National Central University, Jhongli City, *Taoyuan County, 32001, Taiwan*

[6]Univ Rennes, ISCR–UMR 6226, CNRS, F-35000 Rennes, France

[7]Institut Jean Lamour, UMR 7198 CNRS – Université de Lorraine, 54011 Nancy, France

*Corresponding authors: xingchen.shen@ensicaen.fr, emmanuel.guilmeau@ensicaen.fr

**Abstract**

Mixed-anion compounds, which incorporate multiple types of anions into materials, displays tailored crystal structures and physical/chemical properties, garnering immense interests in various applications such as batteries, catalysis, photovoltaics, and thermoelectrics. However, detailed studies regarding correlations between crystal structure, chemical bonding, and thermal/vibrational properties are rare for these compounds, which limits the exploration of mixed-anion compounds for associated thermal applications. In this work, we investigate the lattice dynamics and thermal transport properties of the metal chalcohalides, $CuBiSCl_2$. A high-purity polycrystalline $CuBiSCl_2$ sample, successfully synthesized via modified solid-state synthetic method, exhibits a low lattice thermal conductivity ($\kappa_L$) of 0.9-0.6 W m$^{-1}$ K$^{-1}$ from 300 to 573 K. By combining various experimental techniques including 3D electron diffraction with theoretical calculations, we elucidate the origin of low $\kappa_L$ in $CuBiSCl_2$. The stereo-chemical activity of the $6s^2$ lone pair of $Bi^{3+}$ favors an asymmetric environment with neighboring anions involving both short and long bond lengths. This particularity often implies weak bonding, low structure dimensionality, and strong anharmonicity, leading to low $\kappa_L$. In addition, the strong two-fold linear S-Cu-S coordination with weak Cu⋯Cl interactions induces large anisotropic vibration of Cu or structural disorder, which enables strong phonon-phonon scattering and decreases $\kappa_L$. The investigations into lattice dynamics and thermal transport properties of $CuBiSCl_2$ broadens the scope of the existing mixed-anion compounds suitable for the associated thermal applications, offering a new avenue for the search of low thermal conductivity materials in low-cost mixed-anion compounds.

**Introduction**

The emergence of mixed-anion compounds provides versatility in designing novel materials with unique atomic structures and chemical/physical properties. The combination of different anions within the same crystal structure in mixed-anion compounds, such as oxyhalides,[1] oxychalcogenides,[2] and chalcohalides,[3] offers a wide range of tuneable properties, including optical, magnetic, and electronic properties.[4] These properties can be tailored through the different anionic characteristics of the mixed anions, including charge, ionic radius, electronegativity, and polarizability.[4] Currently, mixed-anion compounds are largely investigated in various fields, such as catalysis,[5] battery electrodes,[6] superconductivity,[7] and optical applications.[8] For instance, oxynitride, $CaTaO_2N$ and $LaTaON_2$ perovskites present tuneable colours for optical applications stemming from their narrowing band gap governed by the mixed anion with the lowest electronegativity.[8]

In addition to their attractive physical properties utilized in optical applications, mixed-anion compounds also hold promise as low thermal conductivity materials.[9] These materials can effectively prevent heat transfer, playing a crucial role in applications such as thermoelectrics,[10] thermal barrier,[11] and heat management.[9d] For example, the mixed anion compound BiCuSeO is an excellent moderate-temperature thermoelectric material due to its low thermal conductivity (0.62-1.28 W m$^{-1}$ K$^{-1}$ at 300 K).[9b, 9c] It adopts the ZrSiCuAs-type structure with the *P4/nmm* space group, and consists of conductive $(Cu_2Se_2)^{2-}$ layers alternately stacked with insulating $(Bi_2O_2)^{2+}$ layers. Prior investigations have suggested that the low thermal conductivity is attributed to the large anharmonicity driven by the presence of lone pair of $Bi^{3+}$[12] and Cu-dominated low-energy vibrations.[13] Recently, an extremely low thermal conductivity of 0.1 W m$^{-1}$ K$^{-1}$ was reported in the mixed anion layered compound $Bi_4O_4SeCl_2$.[9a] This record value, measured in the stacking direction of the structure, stems from the suppressed phonon group velocity through the manipulation of the spatial arrangement of distinct interfaces.[9a]

Despite the low thermal conductivity of BiCuSeO and $Bi_4O_4SeCl_2$, their widespread applications are impeded by the toxicity and scarcity associated with Se. Hence, there is an increasing interest in exploring cost-effective and environmentally friendly materials, exemplified by Cu-based oxy-sulphides/chloro-sulphide. Recently, the novel mixed chalcohalides $CuBiSCl_2$ has emerged as a promising candidate for photovoltaic applications.[14] However, a comprehensive investigation into the interplay among crystal structure, lattice dynamics, and thermal conductivities is lacking. Particularly, the reported $CuBiSCl_2$ sample exhibited a notable presence of impurities, BiOCl,[14] which unavoidably affects its intrinsic chemical/physical properties.

In this study, we thereby intend to make phase pure $CuBiSCl_2$ compound and investigate the relationships between the crystal structure, lattice dynamics, and thermal transport properties. Our thorough structural analysis, in combination with electron localization function (ELF) and calculated $\kappa_L$, suggests that the stereo-chemical activity of lone pair of $Bi^{3+}$ results in weak bonding, low structure dimensionality, and strong anharmonicity present in $CuBiSCl_2$ sample. Additionally, we suggest that the weak bonding between Cu and Cl favors either large anisotropic atomic Cu vibration or structural disorder and creates an antibonding state near Fermi level ($E_F$). The combination of heat capacity fitting and phonon dispersion analysis further unveils the existence of Bi-/Cu-associated low-energy optical phonons, enabling pronounced anharmonic interactions between optical and acoustic modes. Consequently, the lone pair induced 1D character and weak Cu···Cl interactions governs the low thermal conductivity in $CuBiSCl_2$. Therefore, our study provides an in-depth insight into the link between the crystal structure and the thermal transport properties in mixed-anion metal chalcohalides, $CuBiSCl_2$, and opens an avenue to explore cost-effective and environmentally friendly mixed-anion compounds for thermal management applications.

**Results and Discussion**

*Crystal structure description*

The *Cmcm* structure symmetry of the mixed anion metal chalcohalides $CuBiSCl_2$ has previously been described[15] in terms of $CuS_2Cl_4$ octahedrons and $BiS_2Cl_6$ polyhedrons (**Fig. 1a**) *i.e.* adopting the $UFeS_3$ structure type.[16] According to this description, the structure consists of (010) layers of $CuS_2Cl_4$ octahedrons sharing their S apices along *c*-axis and their Cl-Cl edges along *a*-axis. Those $[CuSCl_2]^{3-}$ layers are interconnected through (010) $Bi^{3+}$ layers sharing edges with the $BiS_2Cl_6$ polyhedrons. The analysis of the chemical bonding in the $CuS_2Cl_4$ octahedrons (**Fig. 1b**) shows one anomaly, that is the four Cu-Cl distances (2.79 Å) in the basal plane of the octahedrons are much larger than the two apical Cu-S distances (2.29 Å) in spite of the similar size of the $Cl^-$ and $S^{2-}$ anions. One indeed observes that the Cu-Cl distance is significantly larger than the sum of the ionic radii of $Cu^+$ and $Cl^-$.[17] This clearly suggests a high anisotropy of the chemical bonding of $Cu^+$ in this structure leading to a strong two-fold linear S-Cu-S coordination alike in oxides, while the four equatorial Cu···Cl interactions (2.79 Å) are much weaker with respect to CuCl[18] which exhibits Cu-Cl distances of 2.34-2.38 Å. Importantly, the $Bi^{3+}$ cations are strongly off centred in the $BiS_2Cl_6$ polyhedrons (**Fig. 1c**) leading to two short Bi-S bonds of 2.65 Å and two short Bi-Cl Bonds of 2.76 Å while the four other Bi-Cl distances of 3.21 Å can only be considered as giving weak Bi···Cl interactions.

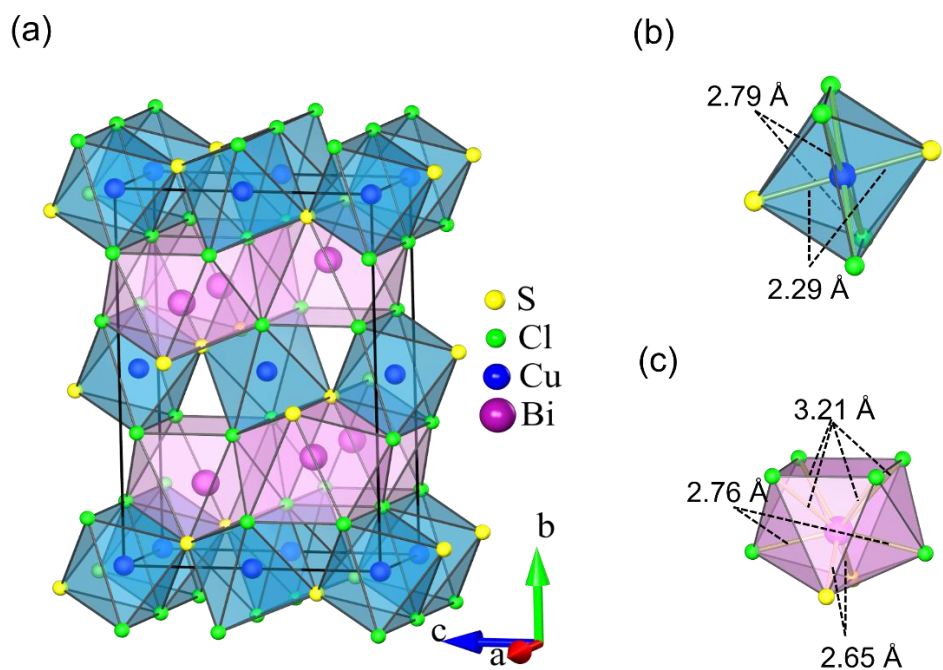

**Fig.1** (a) Polyhedral representation of the $CuBiSCl_2$ structure built up of $CuS_2Cl_4$ octahedrons (turquoise coloured) and $BiS_2Cl_6$ polyhedrons (purple coloured). S, Cl, Cu, and Bi atoms are represented by yellow, green, blue, and purple spheres, respectively. Chemical bonds in $CuBiSCl_2$: (b) $CuS_2Cl_4$ octahedron (c) $BiS_2Cl_6$ polyhedron.

In fact, the coordination of $Bi^{3+}$ is very asymmetric and should be regarded as a $BiS_2Cl_2L$ trigonal bipyramid (**Fig. 2a**) induced by the stereo-chemical activity of the $6s^2$ lone pair (L) of $Bi^{3+}$. These observations demonstrate that the aforementioned polyhedral description is not suitable to understand the nature and especially the anisotropy of the chemical bonding in this compound. An alternative must be considered, based on the fact that the structure of $CuBiSCl_2$ is essentially governed by the stereo-chemical activity of the $6s^2$ $Bi^{3+}$ lone pair. It can be described as the assemblage of $[BiSCl_2L]$ chains of corner-shared $BiS_2Cl_2L$ trigonal bipyramids, running along *a*-axis (**Fig. 2b**). These chains are isolated one from one other but are interconnected through $Cu^+$, which forms a link between two pyramids via S-Cu-S linear bonds (**Fig. 2c**). Therefore, the structure of this metal chalcohalides can be described as a built up of (010) $[Cu_2Bi_2S_2Cl_4]$ flexible bilayers containing copper and characterized by a prominent 1D arrangement of their $BiS_2Cl_2L$ trigonal bipyramids. Note that the Cu···Cl "bonds" (**Fig. 2d**) forming the octahedron $CuS_2Cl_4$ reinforce the stability of those bilayers (**Fig. 2c**) but to a much lesser degree due to the abnormally large Cu···Cl distance of 2.79 Å. Importantly, we observe that the cohesion of the structure along *a* is ensured by weak Bi···Cl interactions of 3.21 Å between two successive bilayers (green dashed lines **Fig. 2c**). Thus, this analysis shows two important characteristics: (i) the structure of this phase consists of $[Cu_2Bi_2S_2Cl_4]$ bilayers, which are made flexible by the fact that their cohesion is only ensured by two-fold coordinated copper; (ii) each bilayer exhibits a prominent one-dimensional character, which is induced by the ability of the bismuth lone pair to generate $[BiSCl_2L]$ chains. The latter are similar to $[BiS_3L]$ chains recently described for the $Pb_mBi_2S_{3+m}$ sulphides,[19] which were shown to exhibit ultralow $\kappa_L$.[20]

In this respect, the present metal chalcohalides should exhibit higher thermal conductivity along *a*-axis due to its strong Bi-S/Bi-Cl bonds in that direction and along *c*-axis due to its strong linear S-Cu-S bonds. Lower thermal conductivity should take place along the *b*-axis where the interactions between the chains are weak. Therefore, the strong 1D character of the structure induced by the $Bi^{3+}$ lone pair appears as the driving force for the generation of low thermal conductivity in this compound. Moreover, the abnormally large Cu···Cl distances in the basal plane of the $CuS_2Cl_4$ octahedrons suggest a weak bonding, which might yield a rattling behaviour of the copper atoms or structural disorder (See next paragraph), which also contribute to lowering the thermal conductivity.

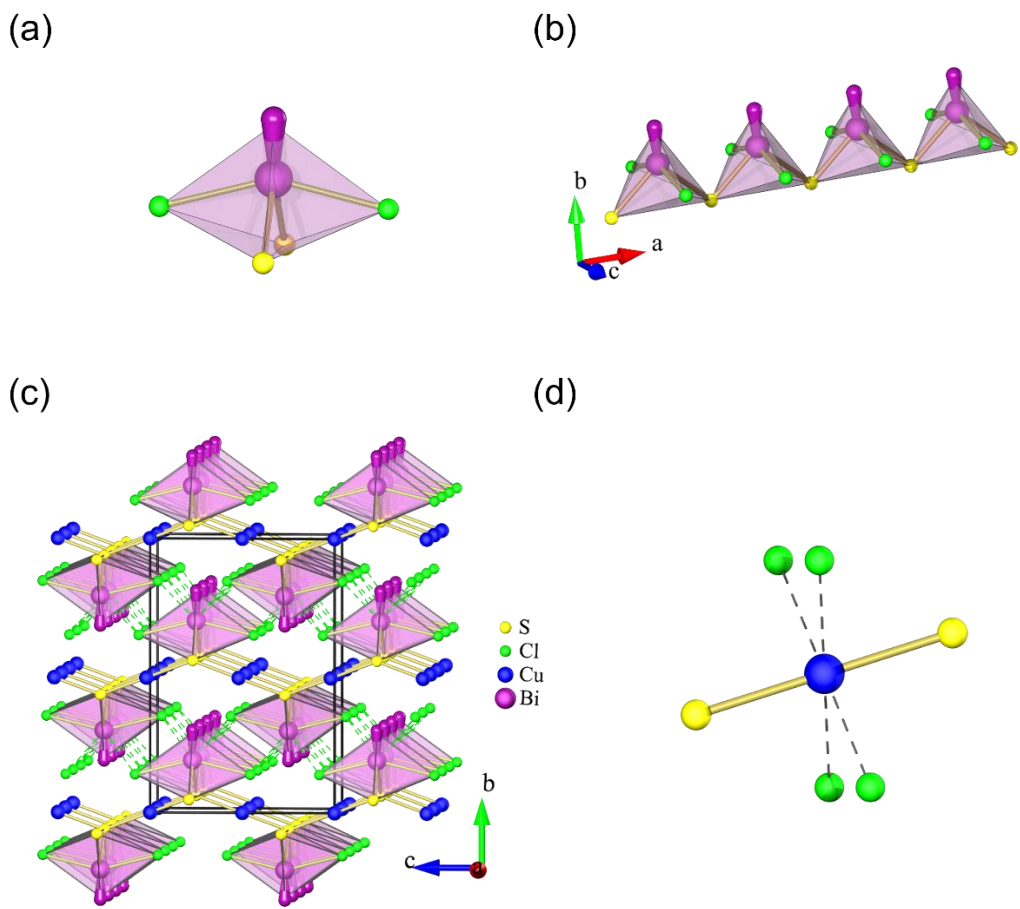

**Fig. 2** (a) BiS$_2$Cl$_2$L trigonal bipyramid showing the lone pair (L) stereo-chemical activity of Bi$^{3+}$. Polyhedral representation of the CuBiSCl$_2$ structure based on the stereo-chemical activity of the Bi$^{3+}$ lone pair. (b) [BiSCl$_2$L] chains of corner-shared BiS$_2$Cl$_2$L trigonal bipyramids, running along *a* (c) view of the structure along *a* showing the (010) [Cu$_2$Bi$_2$S$_2$Cl$_4$] bilayers with copper (blue coloured) in two-fold coordination. The Bi⋯Cl interactions of 3.21 Å, which ensure the cohesion between two successive bilayers, are shown as green dashed lines. d) Coordination of Cu with strong Cu-S bonds and weak Cu⋯Cl interactions of 2.79 Å, shown as black dashed lines.

*Structural analysis: Evidence for large anisotropic atomic displacements of copper*

The powder X-ray diffraction (PXRD) patterns of the synthesized sample at room temperature (RT) confirms its *Cmcm* crystal structure symmetry at 300 K (**Fig. S1**), and our Rietveld refinement suggests the high-purity of our powder sample without any pronounced secondary phase. In contrast to the previous study,[14] where the synthetic samples inevitably had a distinct secondary phase of BiOCl, our modified synthetic method employing excess sulphur avoids the formation of this secondary phase. Moreover, stability measurements were carried out on the CuBiSCl$_2$ powder sample. As illustrated in **Fig. S2a**, prolonged exposure for more than two months ambient air environment led to contamination with a formation of the secondary phase of BiOCl (*P*4/nmm space group). In contrast, maintaining the sample inside the glove box preserved its high-purity quality over time (**Fig. S2b**). Consequently, all the corresponding measurements in this work were conducted on samples securely stored inside the glove box. The corresponding refined atomic parameters, including atomic positions, displacement parameters, and occupancy, are shown in supporting information (**Table S1, SI**). Notably, we found a very large value of the isotropic atomic displacement parameters ($U_{iso}$) for Cu atoms.

To deepen the structural analysis, precession-assisted 3D electron diffraction (3D ED) measurements were conducted on several crystallites of the polycrystalline bulk sample, demonstrating remarkable reproducibility across the dataset. Note that 3D ED[21] is the only single-crystal diffraction method that can be used to study dense polycrystalline materials, as in our sample. The aim here is to obtain a more accurate crystalline structure of our sample, complementary to what is accessible by PXRD. Furthermore, an accurate estimation of the atomic displacement parameters (ADPs) for CuBiSCl$_2$ was possible on the basis of 3D ED data (**Table 1**). This analysis highlighted large anisotropic ADPs for Cu atom, specifically $U_{11}$=0.076 Å$^2$ and $U_{22}$=0.083 Å$^2$ (**Table 1**), represented as blue ellipsoids in **Fig. 3a** and **Fig. S3a**. This characteristic aligns perfectly with the structural findings reported by M. Ruck et al.[15a] using single-crystal X-ray diffraction analysis. This phenomenon appears thus to be an intrinsic characteristic of the CuBiSCl$_2$ rather than a peculiarity of our sample, in relation with the large Cu⋯Cl distances in the basal plane of the CuS$_2$Cl$_4$ octahedrons. Such large ADP values can be attributed to the vibration motion (rattling) of Cu. However, recent studies have found that structural static or dynamic disorder in the form of atomic site splitting also result in large ADP values. This latter phenomenon has been documented in various compounds, including clathrate Sr$_8$Ga$_{16}$Ge$_{30}$,[22] Mo-based cluster Ag$_2$Tl$_2$Mo$_9$Se$_{11}$,[23] Cu$_4$Sn$_7$S$_{16}$,[24] and Cu-Bi-(S, Se)-Cl compounds.[25] Hence, in the present study, an alternative model incorporating a structural disorder was developed using our 3D ED data (**Table 2, Fig. S3b**). However, based on the refinement estimators only, it is not feasible to distinguish between the two models, *i.e.* "rattling" or "splitting" model. This is where chemical bounding considerations come into play.

**Table 1.** CuBiSCl$_2$ structural parameters obtained from single crystal 3D ED data. In this description, similar to the one proposed by Ruck et al.,[15a] large U$_{11}$ and U$_{22}$ values are obtained when modelling ADPs of the Cu atomic position.

| Space Group *Cmcm*, $a$ = 3.9785(7) Å, $b$ = 12.7631(24) Å, $c$ = 8.6045(14) Å |
|---|

| R(obs)=7.32, R(all)= 7.34, wR(all)=17.69, GoF(obs)=5.31 |
|---|

| measured / observed [I>3σ(I)] reflections=3692 / 3656, 125 refined parameters |
|---|

| g$_{max}$ (Å$^{-1}$)=1.6, Sg$_{max}$ (Å$^{-1}$)=0.01, RSg$_{max}$=0.66, steps=256 |
|---|

| Atom | x | y | z | Ueq (Å$^2$) | Occ. |
|---|---|---|---|---|---|
| Cu | 0.5 | 0.5 | 0.5 | 0.065(2) | 1 |
| Bi | 0 | 0.29962(6) | 0.25 | 0.0164(2) | 1 |
| S | 0.5 | 0.4372(2) | 0.25 | 0.0094(6) | 1 |
| Cl | 0 | 0.3525(2) | 0.5622(2) | 0.0135(5) | 1 |

| ADP anisotropic parameters (Å$^2$) | | | | | |
|---|---|---|---|---|---|
| Atom | U11 | U22 | U33 | U12 | U13 | U23 |
| Cu | 0.076(3) | 0.083(2) | 0.037(2) | 0 | 0 | -0.043(2) |
| Bi | 0.0119(4) | 0.0156(4) | 0.0217(4) | 0 | 0 | 0 |
| S | 0.006(2) | 0.013(1) | 0.0091(9) | 0 | 0 | 0 |
| Cl | 0.016(1) | 0.0149(9) | 0.0092(7) | 0 | 0 | -0.0014(5) |

**Table 2.** CuBiSCl$_2$ structural parameters obtained from single crystal 3D ED data. In this alternative description, a structural disorder is assumed for the Cu atomic position. The ADPs for the Cu1a and Cu1b positions were constrained to be identical and refined considering an isotropic value.

| Space Group *Cmcm*, $a$ = 3.9785(7) Å, $b$ = 12.7631(24) Å, $c$ = 8.6045(14) Å |
|---|

| R(obs)=7.36, R(all)= 7.37, wR(all)=17.79, GoF(obs)=5.34 | | | | | |
|---|---|---|---|---|---|
| measured / observed [I>3σ(I)] reflections=3692 / 3656, 125 refined parameters | | | | | |
| $g_{max}$ (Å$^{−1}$)=1.6, $Sg_{max}$ (Å$^{−1}$)=0.01, $RSg_{max}$=0.66, steps=256 | | | | | |
| **Atom** | **x** | **y** | **z** | **Uiso / Ueq (Å$^2$)** | **Occ.** |
| **Cu1a** | 0.5 | 0.4758(4) | 0.5221(5) | 0.0161(7) | 0.5 |
| **Cu1b** | 0.428(2) | 0.5 | 0.5 | 0.0161(7) | 0.5 |
| **Bi** | 0 | 0.2996(1) | 0.25 | 0.0163(2) | 1 |
| **S** | 0.5 | 0.4372(2) | 0.25 | 0.0100(6) | 1 |
| **Cl** | 0 | 0.3526(2) | 0.56201(16) | 0.0139(5) | 1 |

| ADP anisotropic parameters (Å$^2$) | | | | | |
|---|---|---|---|---|---|
| **Atom** | **U11** | **U22** | **U33** | **U12** | **U13** | **U23** |
| Bi | 0.0117(4) | 0.0157(4) | 0.0217(4) | 0 | 0 | 0 |
| S | 0.006(2) | 0.015(1) | 0.0093(9) | 0 | 0 | 0 |
| Cl | 0.016(1) | 0.0160(9) | 0.0092(7) | 0 | 0 | -0.0015(5) |

To better understand the origin of the large anisotropic ADPs of Cu, we also calculated the electron densities of CuBiSCl$_2$ based on the "rattling" or "splitting" mode, which are obtained by the Maximum Entropy Method (MEM, **Fig. 3a**, See details in SI).[26] The electron densities around Cu are very different from the expected ellipsoid and resemble more to a rounded square prism (**Fig. 3b**). It is interesting to note that energy landscape Bond Valence calculations,[27] independent from the obtained structural factors, shows a square shape minimum energy landscape (**Fig. 3c**). This suggests that Cu may not be present in the 4a site (*uvw* = 0,0,0), but could be statistically distributed between four sites transversal to the original Cu-S bond. For such reason, the 3D ED data have been eventually fitted using a split model. In such approach, the distance between Cu and Cl reduces significantly with a minor increase in the Cu-S bond length. The resulting coordination polyhedron for Cu$^+$ is a strongly distorted CuS$_2$Cl$_2$ tetrahedron that statistically reinforces the bilayer cohesion forming Cl-Cu-Cl bridges between the BiS$_2$Cl$_2$L trigonal bipyramid polyhedra in the same layer. Note that energy landscape Bond Valence calculations suggest a relatively flat energy profile (~ 0.07 eV) that would allow also for dynamic disorder at RT. The statistical participation of Cl to the Cu-Cl bond could also explain the larger ADP parameter with respect to the S atoms, which has similar atomic weight. Finally, it is worth noticing from MEM that the deviation from harmonic description of Bi atoms can be, probably ascribed to the presence of the lone pair inducing anharmonic vibrations. The one-dimensional double-well potential energy plot (**Fig. S4**) obtained by mapping the unstable phonon mode of the high-symmetry structure (*Cmcm*) supports the split model of Cu atoms. The presence of multiple energy minimum signify that Cu atoms prefer the low-symmetry positions of these minima than the high-symmetry *4a* site which corresponds to the saddle point (i.e., energy maximum) in the double-well potential energy surface.

Based on our current structural analysis of 3D ED and MEM calculations, we are not able to discriminate unambiguously between rattling and structural disorder to explain the large anisotropic ADPs for $Cu^+$ cations atom.[15b] To accurately determine whether the Cu large anisotropic ADPs has dynamical or static behavior, further complementary experiments are needed, for instance coupling temperature-dependent X-ray diffraction and local structure techniques as EXAFS and Pair Distribution Function as a function of the temperature. In fact, rattling vibration and structural disorder behavior both reflect the rather weak bond strength of Cu-Cl pairs. This weak bonding can contribute to enhance phonon-phonon scattering and lower $\kappa_L$ (See discussion later).

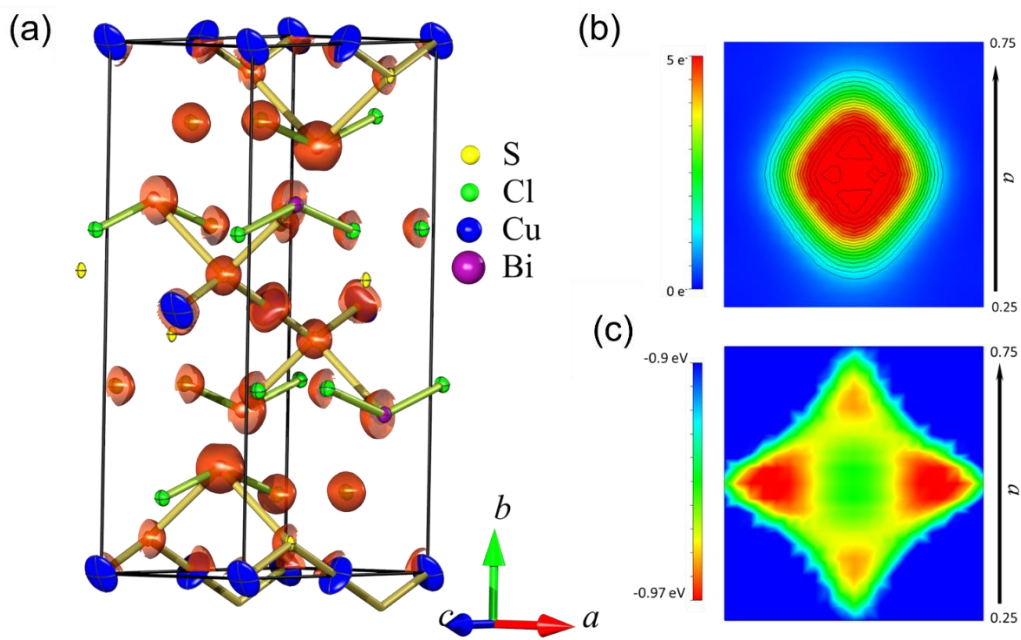

**Fig. 3** (a) CuBiSCl$_2$ structure with anisotropic displacement parameter and electron density as calculated by MEM (isolevel 5 e$^-$). (b) slice of MEM map along the equatorial axes of Cu ADP elipsoid (c) slice of Bond Valence energy landscape calculated for Cu$^{+1}$. Full maps are available in Fig. S5.

The microstructural and chemical analysis of the CuBiSCl$_2$ sample were conducted through scanning electron microscope (SEM). As indicated in the **Fig. S6,** SEM image of a fractured surface present a fine microstructure with a non-uniform size of several micrometers. Besides, the images of the electron energy dispersive spectroscopy (EDS) show a homogeneous distribution of Cu, Bi, S, and Cl elements.

*Chemical bonding and electronic structure*

To qualitatively assess the bonding strengths in CuBiSCl$_2$, the interatomic force constants (IFCs) of all nearest-neighbour cation-anion pairs were calculated. As depicted in **Fig. 4a**, the IFCs of Cu-Cl exhibits an extremely low value of $|\Phi| = 0.19$ eV/Å$^2$ compared to other cation-anion pairs, indicating its very weak bonding. Moreover, the Bi-Cl pairs display the second smallest value of $|\Phi| = 2.72$ eV/Å$^2$, which is still significantly larger than the Cu-Cl bond strength. However, the interaction strength between Bi and Cu with S are much stronger than the rest. Hence, for Bi-S and Cu-S pairs the corresponding IFC values are $|\Phi| = 4.94$ eV/Å$^2$ and $|\Phi| = 3.95$ eV/Å$^2$, respectively. We have calculated the electron localization function (ELF)[28] to identify the presence of lone pairs on Bi$^{3+}$ cations. Indeed, we found the presence of stereo-chemically active lone pair electrons on Bi$^{3+}$ in our analysis, which can be seen as asymmetric charge clouds on Bi$^{3+}$ in the ELF plot (**Fig. 4b**). Hence, the combined IFCs and ELF indicate the existence of weak bonding between Bi-Cl and Cu-Cl pairs in CuBiSCl$_2$, confirming our structural analysis.

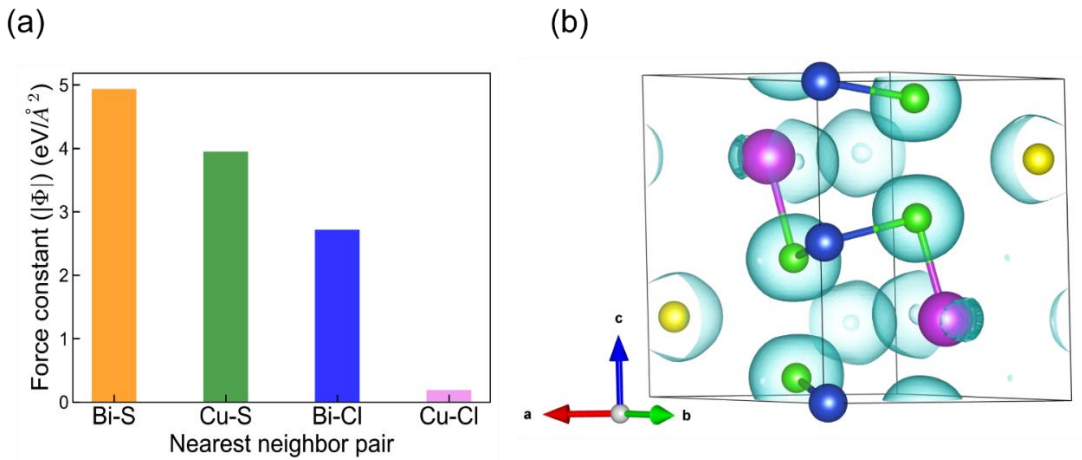

**Fig. 4** (a) Harmonic interatomic force constant (IFC) of the nearest-neighbour Bi-S, Cu-S, Bi-Cl, and Cu-Cl pairs in CuBiSCl$_2$. (b) The electron localization function plot (visualized at an isosurface value of 0.6 e/bohr$^3$) of CuBiSCl$_2$. The asymmetric charge clouds on Bi denote its 6s$^2$ lone pair; Blue, magenta, yellow, and green spheres represent Cu, Bi, S, and Cl atoms, respectively.

To comprehend the nature of the chemical bonding in CuBiSCl$_2$, we performed the crystal orbital Hamilton population (COHP) calculations to analyze pairwise atomic interactions. This approach allows for the quantitative analysis of interaction strengths in terms of bonding, antibonding, and nonbonding states. As shown in **Fig. 5a**, all cation-anion pairs in the mixed anion CuBiSCl$_2$ compound exhibit antibonding states below the Fermi level, suggesting that the mixed anion atomic coordination tends to weaken lattice strength through reducing electron density between different pairs. Particularly, the maximum COHP values of Cu-Cl pair below the Fermi level display significantly larger values (magnitude) of -2.9 eV/cell compared to other pairs, in agreement with the presence of very weak Cu-Cl bonds. **Fig. 5b** illustrates the antibonding states of Cu-Cl/S pairs, providing a visualization of the bonding interaction between Cu (3d) and S/Cl (3p) atoms. The presence of antibonding states near the Fermi level is a signature of low $\kappa_L$, which have been recently observed in metal chalcogenides and metal halides,[29] thus it is expected that the presence of antibonding states in CuBiSCl$_2$ would give low $\kappa_L$. This interaction is further supported by the electronic structure and corresponding electronic density of states presented in **Fig. 5c**, which shows dispersive bands around the conduction band minimum (CBM) dominated by Bi-6p orbitals and relatively less dispersive bands around the valence band maximum (VBM) contributed mostly by Cu-3d orbitals followed by Cl-3$p$ and S-3$p$ orbitals. Moreover, we compare the COHP value of Cu-Cl pair with other *p-d* antibonding states containing inorganic chalcogenides and halide perovskites. As plotted in **Fig. 5d**, the computed COHP value of Cu-Cl in CuBiSCl$_2$ is much larger in absolute value than that of other Cu-based chalcogenides and halide perovskites. The strong antibonding states of Cu-Cl favor the generation of soft lattice and strong anharmonicity, which plays a key role in lowering the $\kappa_L$ in CuBiSCl$_2$. Moreover, the measured optical bandgap of CuBiSCl$_2$ is 1.25 eV from the Kubelka Munk plot (**Fig. S7**), which is consistent with the reported experimental result.[14] The wide band gap nature of CuBiSCl$_2$ suggest that the electronic thermal conductivity contribution is negligible to the total thermal conductivity ($\kappa$) and $\kappa$ is mainly govern by $\kappa_L$.

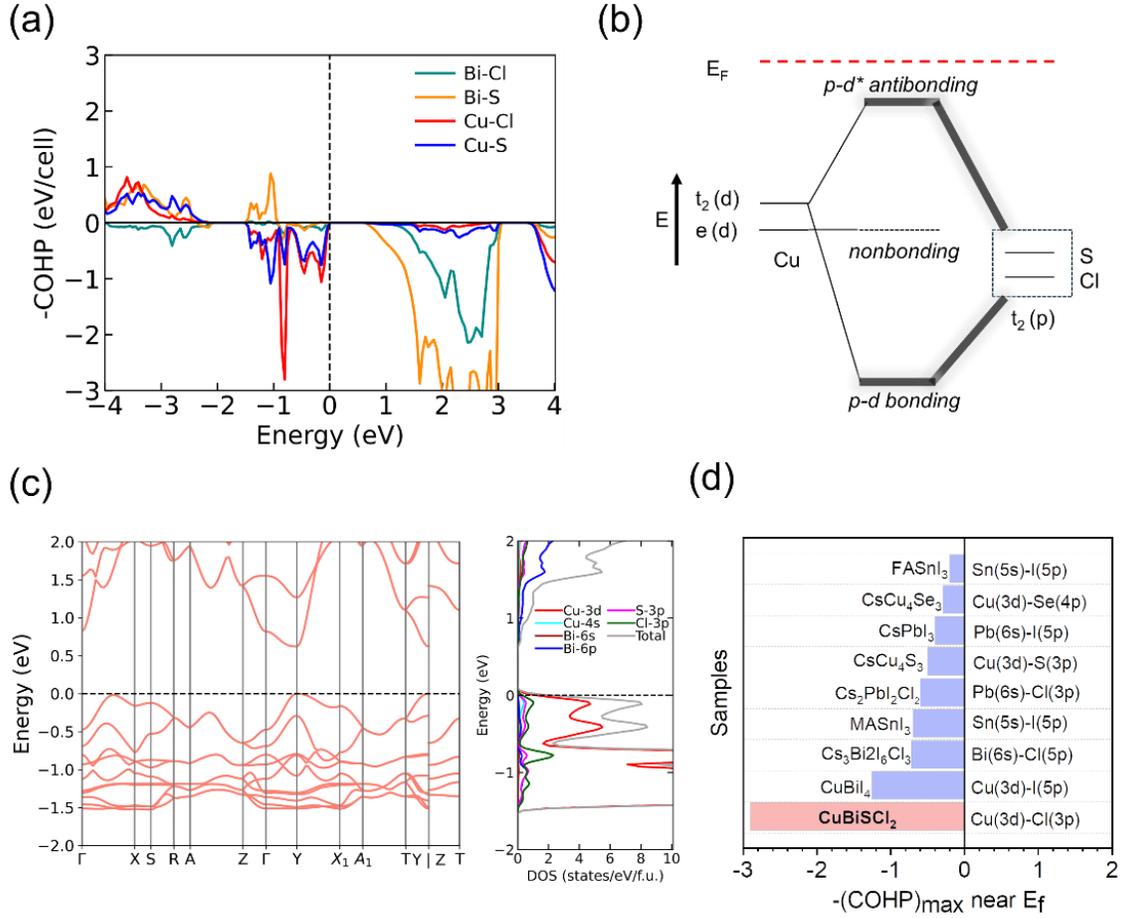

**Fig. 5** (a) Crystal orbital Hamilton population (COHP) plots for the nearest cation-anion pairs in CuBiSCl$_2$. Positive and negative values in the y-axis denote bonding and antibonding states, respectively. (b) Schematic of the molecular energy diagram for CuS$_2$Cl$_4$ octahedrons. (c) The electronic structure and corresponding electronic density of states of CuBiSCl$_2$. (d) Comparing the maximum antibonding value in the COHP plot below the Fermi level ($E_f$) of CuBiSCl$_2$ with that of other inorganic chalcogenides and halide perovskites.[30]

*Lattice dynamics and thermal transport properties*

We have carried out thermal conductivity measurement in the temperature range of 300-573 K (**Fig. 6a**). The room temperature $\kappa_L$ is equal to 0.9 W m$^{-1}$ K$^{-1}$ and decreases following T$^{-1}$ dependency up to the calculated minimum lattice thermal conductivity ($\kappa_{min}$ ~ 0.54 W m$^{-1}$ K$^{-1}$ at 573 K) estimated from the measured sound velocities, which is similar to typical crystalline inorganic materials. We have further compared our data with *state-of-the-art* low thermal conductive materials [31] as shown in **Fig. 6a.** This comparison suggests that our material exhibits intrinsically low $\kappa_L$. We have also calculated $\kappa_L$ of CuBiSCl$_2$ for three axial directions (**Fig. S8**). The calculated $\kappa_L$ is the highest along the *c*-axis, in agreement with the strong linear S-Cu-S bonds in this direction. In contrast, $\kappa_L$ is the lowest along $b$-axis due to the weak bonding interactions of lone pair of $Bi^{3+}$ and Cu-Cl pairs in that direction. Spectral analysis of $\kappa_L$ in **Fig. S9** shows that acoustic phonons with frequencies below 50 cm$^{-1}$ are mainly responsible for lattice thermal conduction along $a$- and $b$-axis. Interestingly, optical phonons between 150 and 200 cm$^{-1}$ also contribute to $\kappa_L$ along the $c$-axis. These phonons involve vibrations weighted mostly by Cl and Cu motions, and to a lesser extent by Bi and S motions (See the corresponding discussion later).

To further understand the origin of low thermal conductivity, we have fitted our experimental data (**Fig. 6b**) using the Debye-Callaway model,[32] given as:

$$\kappa_L(x) = \frac{k_B}{2\pi^2 v_a}\left(\frac{k_B T}{\hbar}\right)^3 \int_0^{\frac{\theta_D}{T}} \frac{x^4 e^x}{\tau_{ph}^{-1}(e^x-1)^2} dx \qquad (1)$$

In this equation, $x$ is the dimensionless parameter $\frac{\hbar\omega}{k_B T}$, where $\omega$ represents the phonon pulsation, $\hbar$ is the reduced Planck constant and $\tau_{ph}$ is the phonon relaxation time. The Debye-Callaway model includes all the possible phonon scattering mechanisms according to the Matthiessen's rule in the forms of various relaxation rates.[33] The scattering rate (inverse of $\tau_{ph}$) can be expressed as

$$\tau_{ph}^{-1} = \tau_B^{-1} + \tau_D^{-1} + \tau_U^{-1} = \frac{v_a}{L} + A\omega^4 + B\omega^2 T e^{\frac{-\theta_D}{mT}} \qquad (2)$$

where $\tau_B, \tau_D$, and $\tau_U$ are the relaxation times associated with the boundary, point defect, and Umklapp scatterings, respectively. All the fitted parameters are listed in **Table S2, SI**. From this fitting, we have calculated the various phonon relaxation times. As demonstrated in **Fig. 6c,** Umklapp scattering (intrinsic) and point-defect scattering (extrinsic) both contributes to lowering $\kappa_L$, as both phonon relaxation times are comparable. It is worth pointing out that very recently a ultralow $\kappa_L$ of 0.27 W m$^{-1}$ K$^{-1}$ at 300 K has been reported by Hawkins *et al.*[15b] for the structurally related CuBiSeCl$_2$. However, the authors could achieve samples with a density of only 88%, which likely contributes to a significant lowering of $\kappa_L$ and hinders a more accurate comparison.

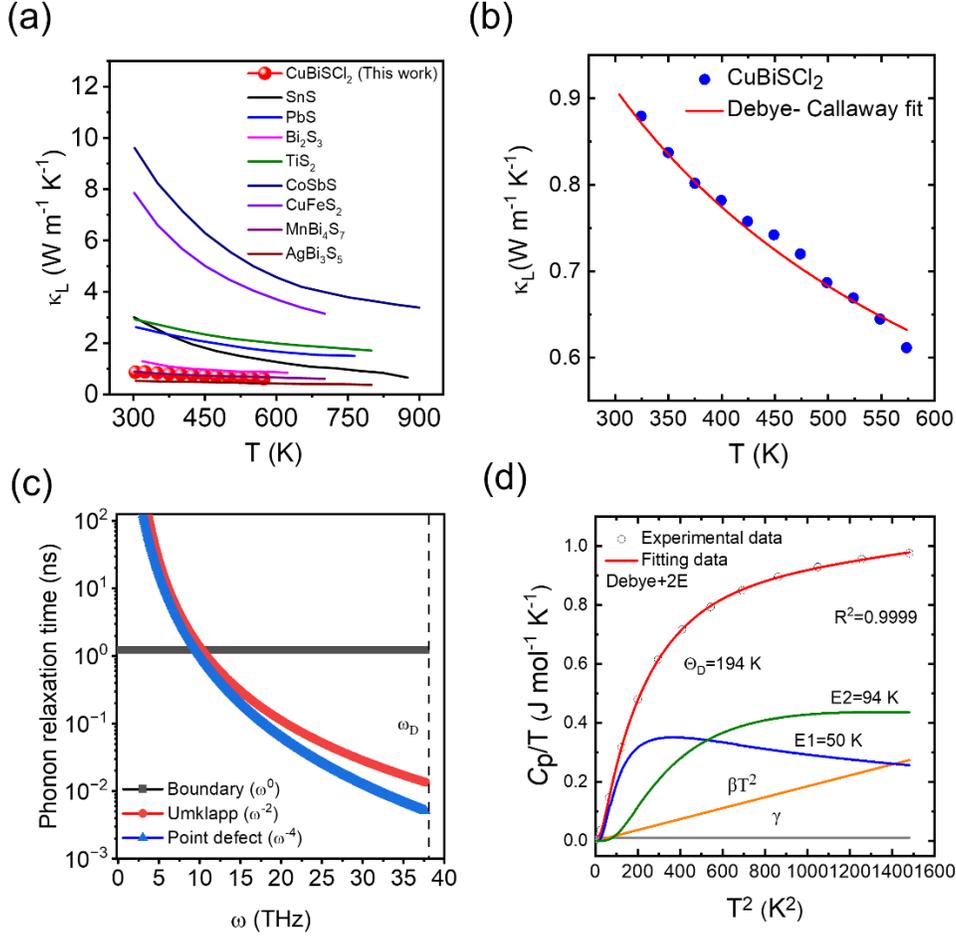

**Fig. 6**(a) Comparison lattice thermal conductivity $\kappa_L$ of CuBiSCl$_2$ with *state-of-the-art* low thermal conductive materials. (b) The measured $\kappa_L$ from 300 to 573 K. The red solid line represents the fitted data using Debye-Callaway model. (c) The calculated phonon relaxation time under various scattering mechanisms, including boundary, point defects, and Umklapp scatterings, $\omega_D$ is Debye frequency. (d) The fitting profile of the experimental data for $C_p/T$ vs. $T^2$ with Debye-Einstein model.

In order to gain deeper insights into the low thermal conductivity observed in CuBiSCl$_2$, we carried out measurement and analysis of low-temperature heat capacity. The $C_p/T$ versus $T^2$ plot was accurately fitted using the Debye-Einstein model, described by the following function:

$$\frac{C_p}{T} = \gamma + \beta T^2 + \sum_n (A_n (\Theta_{E_n})^2 \cdot (T^2)^{-3/2} \cdot \frac{e^{\Theta_{E_n}/T}}{(e^{\Theta_{E_n}/T}-1)^2}) \tag{3}$$

where $\gamma$ is the Sommerfeld coefficient, denoting the electronic contribution to $C_p$, and the term $\beta$ is associated with the lattice contribution and can be determined using the relation $\beta = C(\frac{12\pi^4 N_A k_B}{5}) \cdot (\Theta_D)^{-3}$. Here, $N_A$ and $k_B$ are Avogadro's number and Boltzmann constant, respectively. Additionally, the parameter $C$ is calculated using $C = 1 - \sum_n A_n/3NR$, where $N$ and $R$ are the numbers of atoms per formula unit and the universal gas constant, respectively. The third term represents the lattice contribution arising from Einstein oscillators, with $A_n$ and $\Theta_{E_n}$ denoting the prefactors and Einstein temperatures of the $n^{th}$ Einstein oscillator mode, respectively. As illustrated in **Fig. 6d**, the Debye-Einstein model fitting yields $\Theta_D = 194$ K, $\Theta_{E_1} = 50$ K (34.7 cm$^{-1}$), and $\Theta_{E_2} = 94$ K (65.3 cm$^{-1}$), along with other fitting parameters gathered in **Table S3**. The derived low-energy Einstein modes, $\Theta_{E_1}$ and $\Theta_{E_2}$, are in agreement with the low-energy optical modes observed in the calculated phonon dispersion **(Fig. 7a)**. Furthermore, we obtained the estimated average sound velocity ($v_a$) of ~1819 m s$^{-1}$ using the equation $\Theta_D = \frac{h}{k_B}(\frac{3N}{4\pi V})^{1/3} v_a$, which reasonably aligns with the experimental $v_a$ of 1633 m s$^{-1}$. The measured longitudinal and transverse sound velocities of CuBiSCl$_2$ along with other parameters are shown in **Table S4**. The experimentally obtained value of $v_a$ is in agreement with the estimated value, confirming the accuracy of the $C_p$ fitting.

Phonon dispersion calculations, presented in **Fig. 7a**, reveal the presence of soft acoustic phonon branches whose maximum cut-off frequency is below 50 cm$^{-1}$ along the Γ-X and Γ-Y directions, resulting in low transverse (1857 m s$^{-1}$), longitudinal (3332 m s$^{-1}$), and average (2067 m s$^{-1}$) sound velocities. These lower values of acoustic phonon mode frequencies also lead to low bulk and shear moduli (41 GPa and 22 GPa), indicating the soft lattice governed by the lone pair of Bi$^{3+}$ cations and p-d antibonding states between Cu and Cl atoms. In addition, multiple phonon branches with small dispersions crowd up in the low-frequency region (~ 30 cm$^{-1}$-50 cm$^{-1}$) which can provide an enhanced scattering phase space and hence an increased phonon scattering rates,[34] implying low $\kappa_L$ in CuBiSCl$_2$. As visualized in **Fig. 7b**, the corresponding phonon density of states (phDOS) shows that the low-lying optical phonons are predominantly associated with Cu and Bi atoms.

To quantify the anharmonicity in CuBiSCl$_2$, we calculated the mode Grüneisen parameters ($\gamma_{qv}$), as shown in **Fig. 7c**. The relatively high values (>> 1) of $\gamma_{qv}$ at ~30-50 cm$^{-1}$ confirm the significant anharmonicity associated with the low-energy phonon modes. It has been shown[35] that phonon scattering rates are inversely proportional to the square of $\gamma_{qv}$. Therefore, large $\gamma_{qv}$ for the low-energy phonon modes are synonymous with small phonon lifetimes and hence low $\kappa_L$, in agreement with our experimental data. We have provided eigenvector visualizations of some of the low-frequency phonons that appear at the Γ point (**Figs. 7d-e**). These modes are primarily dominated by vibrations of either Cu and Bi atoms, which govern acoustic-optical phonon scattering and reduce $\kappa_L$.[36]

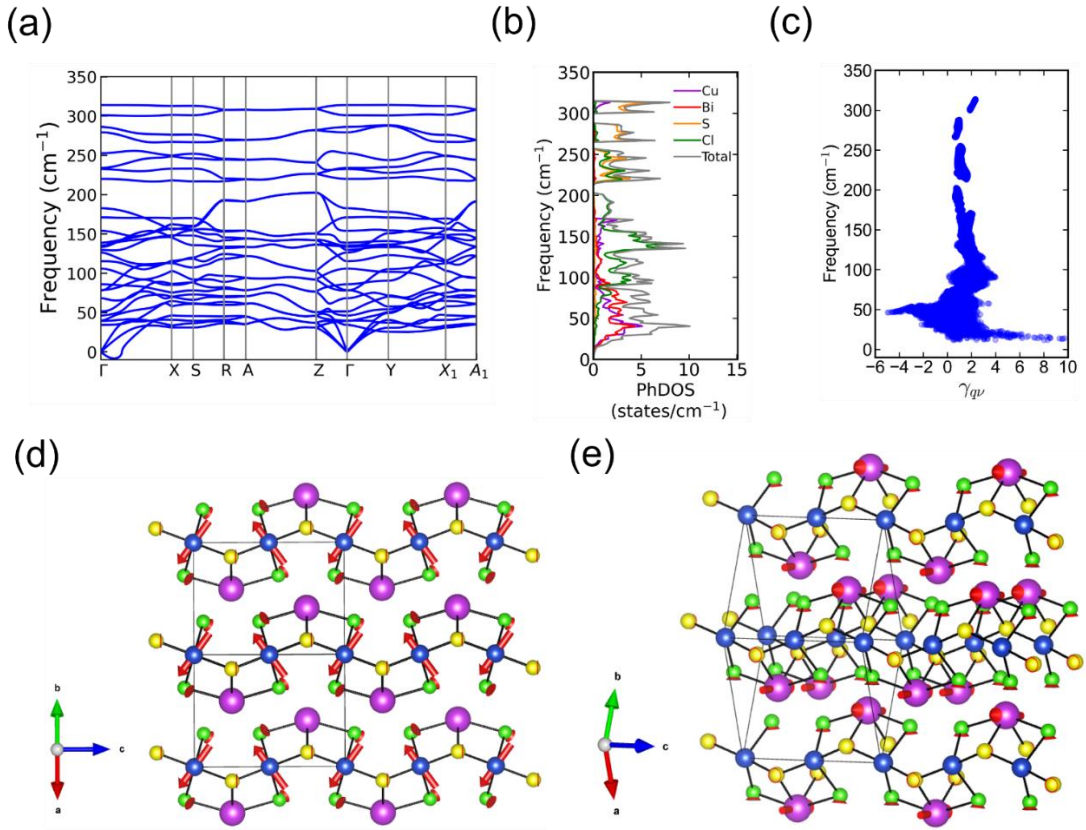

**Fig. 7** The calculated harmonic (a) phonon dispersion, (b) atom-resolved PhDOS, (c) mode Grüneisen parameters ($\gamma_{qv}$) of the phonon modes of $CuBiSCl_2$. (d) Visualizations of two low-frequency phonon modes for (d) 34 cm$^{-1}$, and (e) 40 cm$^{-1}$ at $\Gamma$ point which are dominated by Cu atoms and Bi atoms vibrations, respectively. The black box represents the primitive unit cell. Blue, magenta, yellow, and green spheres represent Cu, Bi, S, and Cl atoms, respectively.

**Conclusions**

In conclusion, we investigated the origin of low $\kappa_L$ in high quality polycrystalline $CuBiSCl_2$. The exceptional character of the metal chalcohalides $CuBiSCl_2$ deals with the great variety of chemical bonding in the structure, which can be classified into two groups: the Bi-S/Bi-Cl and Cu-S strong bonds forming chains of, respectively, $BiS_2Cl_2L$ trigonal bipyramids and $CuS_2$ sticks with inter units Bi⋯Cl and Cu⋯Cl weak interactions. This structural behaviour shows a great similarity with that previously described for the bismuth sulphide $PbBi_2S_4$ in the sense that it is dominated by the stereo-chemical activity of the $6s^2$ lone pair of $Bi^{3+}$. This leads to a prominent 1D structural arrangement of the Bi polyhedral which governs the formation of weak inter-units bonds that is the driving mechanism for the appearance of low $\kappa_L$. Nevertheless, it differs from the sulphides by the fact that the Bi-chains, which give rise to high $\kappa_L$, do not consist of BiS$_3$L tetrahedra but are built up of BiS$_2$Cl$_2$L trigonal bipyramids. This compound is also remarkable by the coexistence of the lone pair induced 1D character of Bi-S-Cl sub-lattice and Cu-induced 2D character of the mixed Bi$_2$Cu$_2$S$_2$Cl$_4$ bilayers. Copper can also contribute to the low $\kappa_L$ of this compound by considering either a two-fold linear S-Cu-S coordination and weak Cu⋯Cl interactions favouring large anisotropic Cu vibrations or a distorted tetrahedral CuCl$_2$S$_2$ coordination resulting in a structural Cu disorder on split sites.

To unearth the origin of the low $\kappa_L$, we have performed in-depth experimental as well as theoretical analysis. The IFCs and COHP calculations further support the weak bond strength between Bi-Cl and Cu-Cl pairs, by the existence of weak force constant and strong antibonding states of Cu-Cl "bonds". Heat capacity analysis has evidenced low-energy optical phonon modes (E1=34.7 cm$^{-1}$) in agreement with calculated phonon dispersions, which enables strong interaction with acoustic phonon modes. Moreover, our calculations show that the Cu and Bi-dominated low-lying optical modes possess large $\gamma_{qv}$ value, indicative of their significant anharmonicity. Through the combination of experimental and computational approaches, this study provides an in-depth understanding of the relationship between the crystal structure, chemical bonding, lattice dynamics, and low $\kappa_L$ in mixed-anion metal chalcohalides CuBiSCl$_2$. These exciting anharmonic phonon properties giving rise to very low $\kappa_L$ evidence the rich interplay between structural chemistry and thermal/vibrational properties in this class of compounds, which will be useful for future exploration of other mixed-anion compounds and their possible use in diverse thermal functional applications.

**Experimental Section**

*Synthesis*

The stoichiometric amounts of the starting precursors CuCl (powder, 99%), BiCl$_3$ (powder, 99.9%), Bi (powder, 99.5%), with a 10% excess S (powder, 99.5%) were weighted and grinded into fine powders inside of an Ar-filled glove box. The fine powders were cold-pressed into pellets and loaded into a carbon coated silica tube, which was then evacuated and sealed at a vacuum of ~10$^{-3}$ Pa. The sealed tube was heated to 703 K within 10 hours and held at this temperature for another 24 hours. Subsequently, it was slowly cooled down to 473 K, followed by annealing at this temperature for 48 hours before turning off the furnace. The obtained ingot was grinded into a fine powder and loaded into a 10 mm graphite die, followed by densification using spark plasma sintering (SPS) at 603 K for 5 minutes under a pressure of 100 MPa. The corresponding SPS-ed sample was verified as highly densified pellet with a 97 % of the theoretical value.

*Powder X-ray Diffraction and Stability measurement*

High-resolution PXRD data of the synthesized SPS-ed powder were collected at 300 K employing a PANalytical XPert2 system with Cu Kα1/ Kα2 radiation (λ1=1.540598 Å, λ2=1.544426 Å). Rietveld refinement was performed using FullProf[37] and WinPlotr[38] software packages starting from the structural model reported by Ruck *et al*.[15a] Zero-point shift, unit cell, peak shape, and asymmetry parameters were refined, as well as atomic coordinates and Biso values of each atoms. Finally, a weak Lorentzian component was refined to correct some broadening effect related to anisotropic particle size (i.e. size model 4 implemented in FullProf). The stability of the synthesized SPS-ed powder was assessed by placing it in ambient air and keeping it inside of glove box for several weeks, respectively, while conducting PXRD testing at regular intervals.

*3D Electron Diffraction*

Precession-assisted 3D electron diffraction (ED) datasets were acquired utilizing a JEOL F200 transmission electron microscope operating at 200 kV. The microscope was equipped with an ASI Cheetah M3 detector and a Nanomegas Digistar precession unit. A powder sample, obtained from a $CuBiSCl_2$ pellet, underwent further grinding in ethanol using an agate mortar. Subsequently, drops of the resulting solution were deposited onto a holey carbon membrane supported by a Cu meshed grid. Precession-assisted electron diffraction (PED) patterns were captured using the Instamatic program,[39] employing a precession angle of 1.25° and a tilt step of approximately 1° between each PED pattern. To ensure the robustness of the results, a total of six $CuBiSCl_2$ crystals were utilized. Data processing was carried out using PETS 2.0,[40] and structure refinements were conducted in Jana2020,[41] accounting for both dynamical diffraction and precession effects. Details regarding data reduction and the outcomes of dynamical structure refinement are provided for one selected dataset in Tables 1 and 2.

*Maximum Entropy Method and Bond Valence Energy Landscape Calculations*

Phased intensity coming from Jana2020 refinement has been analysed using MEM as implemented in Dysnomia code[42] in order to obtain electron-density distributions. Due the dynamic refinement, several reflection were present more than one time the input file for MEM, in such case their values has been averaged and the error was estimated on the base of the standard deviation of the set. Bond Valence energy landscape has been calculated using as input file the structure resulting from dynamic refinement using BondStr code.[43]

*Heat Capacity and Thermal Conductivity Measurements*

The high-temperature thermal conductivity ($\kappa$) was estimated utilizing the formula $\kappa = \rho C_p d$. The thermal diffusivity (*d*) was measured employing a Netzsch LFA 457 laser flash system under a nitrogen atmosphere from 300 to 573 K. $C_p$ measurements were conducted in the temperature range of 2 to 40 K using a conventional relaxation method with the dedicated $^4$He option of the PPMS. The theoretical $C_p$ above RT for CuBiSCl$_2$ is 0.33 J g$^{-1}$ K$^{-1}$ under Dulong−Petit approximation, and the density ($\rho$) was determined via the Archimedes method. $\kappa$ has an estimated measurement uncertainty of 11%.[44]

*Sound Velocity Measurements*

The longitudinal and transverse sound velocities at 300 K were measured using the pulse-echo method. A small quantity of grease was applied to ensure effective contact between the sample and the piezoelectric transducers.

*Optical Measurements*

The UV-vis-NIR diffuse reflectance spectra were recorded with a V-770 JASCO spectrophotometer equipped with an integrated reflectance sphere accessory. The reflectance measurements were made in a wavelength range of 200 to 2500 nm with a 1 nm step on. The optical band gaps ($E_g$) were determined *via* the absorption/back scattering ratio (K/S) calculated from the raw reflectance data using the F(R) = K/S = (1 − R)$^2$/(2R) Kubelka−Munk transform.

*First-principles Calculations*

We performed first-principles calculations based on density functional theory (DFT) calculations using the Vienna Ab-initio Simulation Package (VASP)[45] and the projector augmented wave (PAW)[46] method. We utilized the PBE generalized gradient approximation (GGA)[47] to treat the exchange-correlation energy of the electrons and used a kinetic energy cut-off of 520 eV. Brillouin zone integrations were performed with k-point mesh of $10 \times 10 \times 8$ for the relaxation of the cell parameters as well as for the static calculations. We included the van der Waals interactions in the calculation with Grimme's correction.[48] The optimized lattice parameters (a = 4.055 Å, b = 12.588 Å, c = 8.462 Å) agree well with the corresponding experimental values (a = 3.966 Å, b = 12.811 Å, c = 8.602 Å). We calculated the harmonic phonon dispersion of $CuBiSCl_2$ using the supercell method in Phonopy,[49] where the displaced configurations were generated using 2×2×2 supercell (containing 80 atoms) of the primitive unit cell (with 10 atoms). For phonon calculations, we used 5×5×4 mesh of k-points while calculating the total energy and forces. The calculated phonon dispersion exhibited several unstable phonon modes that span almost the entire Brillouin zone. We examined some of these unstable phonon modes (with frequencies -38 cm$^{-1}$, -29 cm$^{-1}$, and -15 cm$^{-1}$) by visualizing their eigenvectors at Γ point (**Figs. S10a-c** in SI), which reveal that Cu atoms' vibrations dominate them. Next, we took the strongest unstable phonon mode (-38 cm$^{-1}$) at Γ point and calculated its potential energy surface, which shows a double-well structure (**Fig. S11** in SI). However, the depth of the well is 5 meV/cell, which is small, signifying relatively weak nature of the unstable phonon mode. We then took the structure, which stays at the minimum of the potential energy well, optimized the structure which lowers its symmetry (*Ama2*) from the parent high-symmetry phase (*Cmcm*). We calculated the phonon dispersion of this reoptimized structure following the same procedure before. In this newly calculated phonon dispersion, all unstable phonon modes disappear and become stable which is shown in the manuscript. A tiny segment of the first phonon branch shows negative values close to Γ point along the Γ-X line, which we believe is an artifact of the calculations and may disappear if we consider larger supercell in the calculations. However, large supercell will make the calculations computationally very expensive and hence we have not attempted it. The high-symmetry points in the Brillouin zone are taken from Setyawan's work.[50] Subsequent calculations are performed taking this lower-symmetry (*Ama2*) dynamically stable structure. We determined the mode Grüneisen parameters of $CuBiSCl_2$ using a finite difference method where we have calculated the phonon frequencies of the compound at two different volumes (1.02×$V_0$ and 0.98×$V_0$, with $V_0$ being the equilibrium unit cell volume) and utilized the formula $\gamma_{qv} = - d\ln(\gamma_{qv})/d\ln(V)$ where $\gamma_{qv}$ and V denote Grüneisen parameter and unit cell volume, respectively. We calculated the bulk (B) and shear (G) moduli of $CuBiSCl_2$ using the elastic tensor obtain from first-principles calculations and utilizing Voigt's formula[51]. The longitudinal ($v_L$) and transverse ($v_T$) sound velocities are obtained using $v_L = \sqrt{\frac{G}{\rho}}$ and $v_T = \sqrt{\frac{(B + \frac{4}{3}G)}{\rho}}$, respectively, where $\rho$ is the density of the compound. The $v_a$ is calculated using the relation: $\frac{3}{v_a^3} = (\frac{1}{v_L^3} + \frac{2}{v_T^3})$. We used the LOBSTER code [52] to analyze the bonding, antibonding and nonbonding interactions between the atoms using the crystal orbital Hamilton population (COHP) method.

Lattice thermal conductivity tensor was calculated using phono3py.[53] Third-order interatomic force constants were derived by introducing atomic displacements of 0.05 Å into the supercells. 3×1×2 supercells of the conventional unit cell, which have cell lengths larger than 12 Å and contains 120 atoms, were used to avoid the interactions of displaced atoms beyond the periodic boundaries. Total number of displacements needed for thermal conductivity calculations is 11012. To reduce computational costs, we employed a kinetic energy cut-off of 420 eV and a 2×2×2 Monkhorst-Pack mesh of $k$-points[54] in the force calculations. Note that we preliminarily confirmed that these conditions have negligible influence on the accuracy of the calculated forces. The lattice thermal conductivity tensor was calculated using Boltzmann transport equation with the single mode relaxation time approximation, using a fine $q$-point mesh of 21×27×27 for the first Brillouin zone of the primitive cell. The calculated values of $\kappa_L$ are 0.29-0.15 W m$^{-1}$ K$^{-1}$ on average from 300 to 573 K (**Fig. S8**), being underestimated compared to the measured values of 0.9-0.6 W m$^{-1}$ K$^{-1}$. This underestimation is probably due to the crystal structure of CuBiSCl$_2$ used in the phonon calculations which was less symmetric (space group *Ama2* with displaced Cu$^+$ ions) than the experimentally observed structure (*Cmcm*). A more symmetric *Cmcm* structure may be stabilized by anharmonic vibrations at finite temperature and increase $\kappa_L$ in experiments.

**Acknowledgement:**

X.S. acknowledges funding from the European Union's Horizon 2020 research and innovation program under the Marie Sklodowska-Curie grant agreement No. 101034329 and the WINNINGNormandy Program supported by the Normandy Region. K.P. acknowledges financial support from the Initiation Grant by Indian Institute of Technology Kanpur. The support and the computational resources provided by HPC2013 and PARAM Sanganak under the National Supercomputing Mission, Government of India are gratefully acknowledged. S.F. was supported by KAKENHI (Grant No. JP22H04914 and JP23K13544) from the Japan Society for the Promotion of Science (JSPS). The computations for lattice thermal conductivity were mainly performed using the Supercomputer Center in the Institute for Solid State Physics, the University of Tokyo.